\documentstyle[aps,epsf,multicol]{revtex}
\begin{document}                                      
\draft
\widetext
\title{Fractional quantum Hall junctions and two-channel Kondo models}
\author{Nancy P. Sandler${}^{1,2}$ and Eduardo Fradkin${}^3$}
\address{
${}^1$ LPTMS UMR 8626 CNRS-Univ. Paris Sud, Bat. 100; 91405 Orsay France  \\
${}^2$ Dept.\ of Physics, Brandeis University, Waltham MA 02454\\
${}^3$ Dept. of Physics, University of Illinois at Urbana-Champaign,
Urbana, IL 61801-3080 USA}
\maketitle
\begin{abstract}
  A mapping between fractional quantum Hall (FQH) junctions and the
  two-channel Kondo model is presented.  We discuss in detail this
  relation for the particular case of a junction of a FQH state at
  $\nu=1/3$ and a normal metal. We show that in the strong coupling
  regime this junction has a non-Fermi liquid fixed point. At this
  fixed point the electron Green's function has a branch cut. Here we also
  find that at this fixed point there is a non-zero value of the
  entropy equal to $S = \frac{1}{2} \ln{2}$.  We construct the space of
  pertubations at the strong coupling fixed point and find that the dimension of the
  tunneling operator is $1/2$. These behaviors are strongly reminiscent of
  the non-Fermi liquid fixed points of a number of quantum impurity
  models, particularly the two-channel Kondo model. However we have found that, in spite
  of these similarities, the Hilbert
  spaces of these two systems are quite different. In particular,
  although in a special limit the Hamiltonians of both systems are the same, their
  Hilbert spaces are not since they are
  determined by physically distinct boundary conditions. As a
  consequence the spectrum of operators in both problems is 
  different.
\end{abstract}


\begin{multicols}{2}
\narrowtext
\newcommand{\beq}{\begin{equation}}
\newcommand{\eeq}{\end{equation}}
\newcommand{\beqa}{\begin{eqnarray}}
\newcommand{\eeqa}{\end{eqnarray}}
\newcommand{\ra}{\rightarrow}
\newcommand{\la}{\leftarrow}     
\section{Introduction}
\label{sec:intro}

In recent years, advances in nanofabrication techniques have made
possible to study in a controlled manner the physics of the two
channel Kondo model (2CK) in real low-dimensional
materials\cite{ref:2CK-experiments,ref:chang}. Systems like quantum point
contact tunnel junctions, quantum dots connected to leads, and quantum
Hall tunnel junctions; have been proposed as good candidates to
exhibit the full range of 2CK physics by an appropriate variation of
the experimental parameters
\cite{ref:furusaki,ref:matveev,ref:yi&kane,ref:kane,ref:Wen1}. 

All of these experiments can be described in terms of the conceptual
setting of the Kondo model, {\it i.e.} considering the effects of a
local interaction of fermions with a quantum impurity. While in the
one-channel Kondo model one deals with a set of effectively
one-dimensional (radial) fermions -in practice a Fermi liquid- coupled
to a point-like quantum mechanical magnetic impurity, in the
multichannel model the situation is complicated either by a band
degeneracy or by an orbital impurity.  The theoretical
description of the systems mentioned above is rather natural in terms
of a 2CK model, as it can be seen, for instance, in the case of
quantum dots: at resonance the dot contains two degenerate levels that
can be associated with the impurity spin states (up and down), while
the two leads can be represented by the channels. When the leads are
modelled by Fermi liquids, this identification is extremely useful in
the analysis of experimental results for the various observed
conductance regimes.

Another set of important and exciting new experiments have been
carried out in similar settings with systems that are best
characterized as Luttinger liquids. Perhaps the cleanest, from the
point of view of the physics, are the tunneling
experiments\cite{ref:chang} in which electrons from an external
three-dimensional reservoir are injected into the edges of a
fractional quantum Hall state. Just as important are recent noise
experiments\cite{ref:noise} which, having made possible the study of
the properties of quasiparticles in FQH states, provided further
support to the Luttinger liquid picture for edge states. Also,
Luttinger physics seems to be the appropriate description for
transport experiments in carbon nanotubes and nanotube
junctions\cite{ref:nanotubes}.

Hence, it is tempting to model these systems also with the
two-channel Kondo model picture, {\it i.e.} two channels locally
interacting with a quantum impurity. For example, in the case of FQH
junctions, the tunneling experiments involve transfer of electrons
from a Fermi liquid external lead to an edge state at a weak tunneling
center, or tunneling of quasiparticles between two edges in a Hall bar
experiment at a constriction. In the case of nanotubes there are
junctions at the endpoints of the tubes or internal junctions due to
kinks in the tube itself. Thus, tunneling centers, constrictions and
kinks play the role of the quantum impurity, while the channels are
not Fermi but Luttinger liquids.

The existence of a connection between quantum impurity problems, (such
as Kondo models), and FQH junctions has been known for some time
particularly since the work of Kane and
Fisher\cite{ref:kanefisher}. As it turns out, the universal properties
of all these models can be understood in terms of a generic
quantum impurity problem. Furthermore, all quantum impurity problems -at
least their universal behavior- can be classified in terms of an
appropriate set of boundary conditions for a suitable
$1+1$-dimensional boundary conformal field
theory\cite{ref:affleck-ludwig}.

Our recent work on quantum Hall tunnel junctions has yielded evidence
for a relation between the properties of the junctions in the strong
coupling regime and some effective 2CK model \cite{ref:andreev}. The
purpose of this paper is to investigate these analogies further.  A
crucial problem discussed below is the relation between the
description of the isolated FQH edges and their Hilbert spaces as
given by the chiral Luttinger liquid model proposed by
Wen\cite{ref:Wen}, and their 2CK model counterparts.  Although chiral
Luttinger liquids are examples of boundary conformal field theories,
the relation with the 2CK model at this level is still far from
obvious since it is not clear if the Hilbert spaces can be mapped into
each other or not. Thus, in order to make possible such a mapping it
is necessary to carry out a detailed comparison between the respective
Hilbert spaces. The junction model that looks like the more suitable
candidate to compare with the 2CK model is the one between a $\nu =
1/3$ FQH state and a normal metal. In fact, analysis of different
theoretical models \cite{ref:kane,ref:andreev}, for this particular
junction seem to indicate that, in the strong coupling regime, it
posses several properties that appear in the 2CK physics like:

\begin{enumerate}
\item
Both systems have Fermi liquid and non-Fermi liquid fixed points.
\item
At their respective non-Fermi liquid fixed points they both have a
vanishing one-body $S$-matrix (albeit of physically different
operators)
\item
They both have a non-integer value for the boundary entropy at their
respective non-Fermi liquid fixed points equal to
$S_{2CK}=\frac{1}{2}\ln 2$. For a general FQH junction of a Laughlin
state the entropy is given by $S_{FQH/NM}=\frac{1}{2}\ln(k+1)$, where
$2k+1 = \nu^{-1}$. Thus, for $k=1$ corresponding to $\nu = 1/3$, the
entropy has the 2CK value\cite{ref:destri,ref:wiegmann}.
\item
In both cases the renormalization group (RG) flow from the non-Fermi
liquid to the Fermi liquid fixed point is induced by a relevant
operator of scaling dimension $1/2$. While in the 2CK, this
perturbation corresponds to a potential breaking the channel symmetry
\cite{ref:affleck,ref:fabrizio,ref:andrei}; in the FQH/NM junction it
corresponds to the tunneling term at the strong coupling fixed point.
\end{enumerate}

Considering all these similarities, it is tempting to conclude that
FQH junctions are in fact equivalent to an effective 2CK system. This
is most remarkable given that both systems actually have different
symmetries \cite{ref:affleck}. It is thus appropriate to inquire how
the two descriptions are actually related. In this paper we carry out
a detailed analysis of the apparent mapping between the 2CK and a $\nu
= 1/3$ FQH/NM junction in the strong coupling regime. The guiding
motivation, is to establish a formal mapping between operator contents
that may be used, in principle, to translate the results obtained in
one model to the other one. Thus, for example, a FQH junction with two
point contacts could be mapped to a two-impurity 2CK problem and
hence, the problem of a FQH 'quantum dot' could be treated by the same
techniques available for the 2CK model. We will show in this paper
that, at the level of their effective Hamiltonians, there is a mapping
between the FQH/NM junction and the flow induced by channel anisotropy
on the 2CK system. In the problem at hand we actually find a mapping
to the Tolouse limit of the two-channel Kondo problem. However, it is
well known that the 2CK system at the Tolouse point flows to the
non-trivial fized point of the isotropic system and thus the aparent
problem with the symmetry is avoided. However, a closer examination of
the two systems reveals that at the level of their Hilbert spaces they
are actually not equivalent. In particular the spectrum of allowed
operators is not the same.

The paper is organized as follows. After briefly reviewing the model
proposed for a FQH/NM junction (Sec.~\ref{sec:model}), we discuss the
$\nu = 1/3$ junction and discuss the analogies with the two-channel
Kondo problem. In Sec.~\ref{sec:dimensions}, we show that the
perturbation produced by a tunneling operator (in the dual picture)
has dimension $1/2$ as in the 2CK case.  In Sec.~\ref{sec:entropy} we
show that the value of the entropy at the non-Fermi liquid fixed point
coincides with the 2CK boundary entropy. In
Sec.~\ref{sec:greenfunction}, we show that at the non-Fermi liquid
fixed point the Green's function for electrons on the normal metal
side of the junction has a branch cut singularity.  Thus, at this
fixed point, the one-body scattering matrix vanishes as it does in the
2CK system.  We sketch the steps in the calculation for the junction
in equilibrium, {\it i.\ e.\/} at $V=0$ (zero voltage), and in
non-equilibrium ($V \neq 0$).  Next, by mapping the electron operator
present on the normal metal side of the junction, we discuss in
Sec.~\ref{sec:electron} the mapping of the FQH/NM junction to the
Toulouse limit of the 2CK problem. Here we show that, in spite of the
close analogies, both models have {\sl different operator content},
{\it i.\ e.\/}, their Hilbert spaces are different.  The discussion of
these results are summarized in Sec.~\ref{sec:conclusions}.

\section{Model of a FQH/NM junction }
\subsection{Model for a FQH junction}
\label{sec:model}

We start this section with a review of the model for a FQH/NM junction
used in Ref.\cite{ref:andreev}. The Lagrangian for the FQH/NM junction
that describes the dynamics on the edge of a FQH liquid, the electron
gas reservoirs, and the tunneling between them at a single
point-contact is:
\begin{equation}
{\cal L} = {\cal L}_{edge} + {\cal L}_{res} + {\cal L}_{tun}\,.
\label{eq:Lag4}
\end{equation}

The dynamics of the edge of the FQH liquid with a Laughlin filling
factor $\nu=\frac{1}{2k+1}$ is described by a free chiral boson
field $\phi_1$ with the Lagrangian \cite{ref:Wen}
\begin{equation}
{\cal L}_{edge} = \frac{1}{4\pi}
\partial_x \phi_1 (\partial_t - \partial_x) \phi_1\,.
\label{eq:Led4}
\end{equation}
where the units have been chosen so that the velocity of both bosons
is $v=1$ (this is consistent for a coupling at a single point in
space).

The operators that create electrons and quasiparticles at the edge of
a Laughlin state with filling factor $\nu$ are given by
\begin{eqnarray}
\psi_{e}  &\propto & \,  \eta_{\rm edge} \; 
:e^{-i\frac{1}{\sqrt{\nu}} \phi_1(x,t)}:
\nonumber \\
\psi_{qp} &\propto & \, \eta_{\rm edge} \; :e^{-i \sqrt{\nu} \phi_1(x,t)}:
\nonumber \\
&&
\label{eq:psiedge4}
\end{eqnarray}
where we introduced the Klein factor $\eta_{\rm edge}$. 

In Eq.(\ref{eq:Lag4}), ${\cal L}_{res}$ describes the dynamics of the
electron gas reservoir. As shown in Ref. \cite{ref:ClaudioEduardo}, a
2D or 3D electron gas can be mapped to a 1D chiral Fermi liquid (FL)
($\nu = 1$) when the tunneling is through a single point-contact. This
1D chiral Fermi liquid is represented by a free chiral boson field
$\phi_2$. ${\cal L}_{res}$ is given by
\begin{equation}
{\cal L}_{res} = \frac{1}{4\pi} \partial_x \phi_2 (\partial_t - \partial_x)
\phi_2\,.
\label{eq:Lres4}
\end{equation}

In this case, the electron operator is given by
\begin{equation}
\psi_{res} \propto \; \eta_{\rm res} \; :e^{-i \phi_2(x,t)}:
\label{eq:el4}
\end{equation}
where $\eta_{\rm res}$ is the corresponding Klein factor for the
electrons on the metal side of the junction.

The tunneling Lagrangian between the FQH system and the reservoir is
\begin{equation}
{\cal L}_{tun} = \Gamma \; \delta (x)\; \eta_{\rm edge}^\dagger \eta_{\rm res}
\; :e^{i[\frac{1}{\sqrt{\nu}} \phi_1(x,t) - \phi_2(x,t)]}:
+{\rm h.\ c.\ }\,,
\label{eq:Ltun4}
\end{equation}
where $\Gamma$ represents the strength of the electron tunneling
amplitude which takes place at a single point in space $x=0$, the
point contact. In Eqs. (\ref{eq:psiedge4})-(\ref{eq:el4}) the Klein
factors insure that an electron on the FQH edge anticommutes with an
electron on the reservoir. The simplest choice\cite{ref:halpern} is to
define them as
\begin{eqnarray}
\eta_{\rm edge}&=&:e^{\displaystyle{i \frac{\pi}{2} Q_{\rm res}}}:
\nonumber \\
\eta_{\rm res} &=&:e^{\displaystyle{-i \frac{\pi}{2} Q_{\rm edge}}}:
\nonumber \\
\end{eqnarray}
which satisfy the correct mutual statistics. Here $Q_{\rm edge}$ and
$Q_{\rm res}$ are the total charge of the FQH edge and the Fermi
liquid reservoir respectively.  Since the total charge $Q_{\rm
edge}+Q_{\rm res}$ is conserved by the tunneling process, the factor
$\eta_{\rm edge}^\dagger \eta_{\rm res}$ in Eq.(\ref{eq:Ltun4}) is a
constant of motion and as such it can be absorbed in the coupling
constant. In Sec.\ \ref{sec:electron} we will find an analogous set of
operators in the context of the two-channel Kondo problem.

In what follows, by analogy with quantum impurity problems, we will
refer to the point contact as the impurity. Notice however that, in
contrast with the impurity as in the 2CK model, the point contact does
not have internal degrees of freedom. In spite of this difference we
still will find a close analogy between the two problems.

An external voltage difference between the two sides of the junction
can be introduced in the model by letting $\Gamma \rightarrow \Gamma
e^{-i \omega_0 t}$, where $\omega_0 = e V/\hbar$. This external
voltage $V$ can be interpreted as the difference between the chemical
potentials of the two systems: $V = \mu_1 - \mu_{\nu}$. In what
follows the tunneling amplitud will be considered to be complex, in
order to deal with the more general non-equilibrium situation. For the
equilibrium calculation we will simply put $V=0$, thus making the
coupling constant $\Gamma$ real.

By a suitable rotation the original Lagrangian ${\cal L}$ can be
mapped into a new one \cite{ref:andreev,ref:ClaudioEduardo}:
\begin{equation}
{\cal L} = {{\cal L}_0}^{'} \;+\;\Gamma\, \delta(x)\, e^{-i \omega_0 t}
e^{i \frac{1}{\sqrt{g'}}[\phi'_1(x,t) - \phi'_2(x,t)]} + h.c.
\label{eq:Lab4}
\end{equation}
where the new fields $\phi'_1$ and $\phi'_2$ have a free dynamics
governed by ${\cal L}_0^{'}$ and $g'$ is an effective Luttinger
parameter (which can also be regarded as an ``effective filling
fraction'') given by:
\begin{equation}
g'^{-1} = \frac{(1+\nu^{-1})}{2}.
\label{eq:g'4}
\end{equation}

The rotation matrix is given by:
\begin{equation}
\left(\matrix{\phi_1'\cr \phi_2'}\right)
=
\left(\matrix{
\cos\theta & \sin\theta\cr
-\sin\theta & \cos\theta}\right)
\left(\matrix{
\phi_1\cr \phi_2}\right)\ ,
\label{eq:rotation}
\end{equation}
where the angle $\theta$ is determined by
\begin{equation}
\cos 2\theta = \frac{2\sqrt{\nu}}{1+\nu} \,\; ,\;
\sin 2\theta = \frac{1-\nu}{1+\nu}
\label{eq:costheta}
\end{equation}

Next, we introduce the fields $\phi_-$ and $\phi_+$  that separate
${\cal L}$ into two decoupled Lagrangians ${\cal L}_+$ and ${\cal L}_-$,
\begin{equation}
\phi_+ = \frac{\phi'_1 + \phi'_2}{\sqrt{2}}\,\; ,\;
\phi_- = \frac{\phi'_1 - \phi'_2}{\sqrt{2}}.
\label{eq:rot2}
\end{equation}

In terms of the $\phi_{\pm}$ fields the total Lagrangian reads:
\begin{equation}
{\cal L} ={\cal L}_0(\phi_+;\phi_-)+\Gamma \;\delta(x) \;e^{-i \omega_0 t} \; e^{\sqrt{\frac{2}{g'}}
\phi_-(x,t)}+h.c. 
\label{eq:lagpm}
\end{equation}
where ${\cal L}_0(\phi_+;\phi_-)$ describes the free dynamics.  The
strong coupling limit of this system is best described in the dual
picture with the dual fields ${\tilde \phi}_{\pm}$. In terms of these
dual fields, the effective Lagrangian has a new Luttinger parameter
${\tilde g}'=1/g'$ and an effective tunneling amplitude ${\tilde
\Gamma}\sim \Gamma^{-{\frac{1+\nu}{2\nu}}}$. The (dual) Lagrangian is
\begin{equation}
{\tilde {\cal L}} = {\tilde {\cal L}}_0(\phi_+;\phi_-)+{\tilde \Gamma} \;\delta(x) \;e^{-i \omega_0 t}\;
e^{i\sqrt{2g'} {\tilde \phi}_-(x,t)}+h.c.
\label{eq:duallag}
\end{equation}

Working exactly at the strong coupling fixed point, the
duality transformation can also be written as follows:
\begin{eqnarray}
\tilde\phi'_1 &=&
\phi'_1 \Theta(-x) + \phi'_2 \Theta(x) \nonumber\\
\tilde\phi'_2 &=&
\phi'_2 \Theta(-x) + \phi'_1 \Theta(x)\ .
\label{eq:dualfi4}
\end{eqnarray}
where $\Theta(x)$ is the step function. The expressions for the dual
fields in terms of the original fields correspond to an effective
change in the basis used to describe the model. This change of basis
can be carried out only at the strong coupling fixed point {\it i.e.}
${\tilde \Gamma} = 0$. 

\section{Fixed Points of the FQH/NM junction}
\label{sec:dimensions}
In this section we analyze the fixed points of the model for a FQH/NM
junction descrbed in the previous section. We begin by calculating the
scaling dimension of the tunneling operator in each fixed point, {\it
i.\ e.\/} $\Gamma = 0$ and $\Gamma \to \infty$.  We follow closely the
procedure developed by Cardy \cite{ref:cardy} to obtain the scaling
dimension for a boundary operator. At the end of this section we
calculate the boundary entropy at the strong coupling fixed point.
\subsection{$\Gamma = 0$ Fixed Point}

\begin{figure}
\vspace{.2cm}
\noindent
\hspace{.8cm}
\epsfxsize=3.0in
\epsfbox{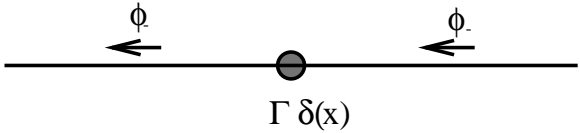}
\vspace{0.9cm}
\caption{Chiral boson field $\phi_-(x,t)$ with a quantum point contact modeled
as an impurity at the origin.}
\label{fig:fig1.fig}
\end{figure}

As stated in Sec.~\ref{sec:model}, the Lagrangian describing the junction
is:
\beqa
{\cal L} &=& \frac{1}{4\pi}
\partial_x \phi_+ (\partial_t - \partial_x) \phi_+ + \frac{1}{4\pi}
\partial_x \phi_- (\partial_t - \partial_x) \phi_- \nonumber \\
&+& 2 \;\Gamma \;\delta(x)\; \cos\Big(\sqrt{\frac{2}{g'}} \phi_-(x,t)\Big)
\label{eq:lag1}
\eeqa
where the field $\phi_-(x,t) = \phi_-(t+ix)$ extends from $x \to -\infty$ to
$x \to \infty$ (see Fig.~(\ref{fig:fig1.fig})). 

\begin{figure}
\vspace{.2cm}
\noindent
\hspace{0.8cm}
\epsfxsize=3.0in
\epsfbox{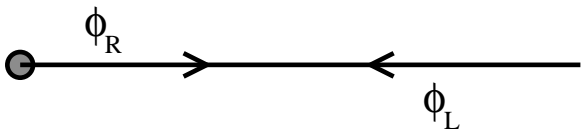}
\vspace{0.9cm}
\caption{Non-chiral theory in terms of the chiral fields $\phi_{R,L}$ with
an impurity in the origin.}
\label{fig:fig2.fig}
\end{figure}

We now proceed to transform the chiral theory into a non-chiral one on the
semi-line by the transformation:
\beqa
\phi_L(x,t) &=& \phi_-(x,t) \nonumber \\
\phi_R(x,t) &=& \phi_-(-x,t)
\label{eq:pilpir}
\eeqa
where the fields $\phi_{R,L}(x,t)$ are defined on the semi-line $x \geq 0$ as 
shown in Fig.~(\ref{fig:fig2.fig})

It is straightforward to show that the sum of the free chiral Lagrangians
in Eq.~(\ref{eq:lag1}) corresponds to a free theory for the non-chiral 
field $\phi$ defined as
\beqa
\phi(x,t) &=& \phi_R(x,t) + \phi_L(x,t) \nonumber \\
{\cal L}_0&=& \frac{1}{8\pi} \int_0^L dx [(\partial_t\phi)^2 - (\partial_x\phi)^2] 
\label{eq:phi}
\eeqa

The tunneling term is transformed in the following way
\beqa
& &{\cal L}_{tun}= 2 \int_{-\infty}^{\infty}\;dx\;\Gamma\;\delta(x)\;\cos\Big(\sqrt{\frac{2}{g'}}\;\phi_-(t+ix)\Big) \nonumber \\
&=& 2 \int_{0}^{\infty}\;dx\;\Gamma\;\delta(x)\;
\Big[\cos\Big(\sqrt{\frac{2}{g'}}\;\phi_L(t+ix)\Big) \nonumber \\
&+&\cos\Big(\sqrt{\frac{2}{g'}}\;\phi_R(t-ix)\Big)\Big]
\label{eq:tunnlag}
\eeqa

By using the fact that $\phi_R(x=0,t) = \phi_L(x=0,t)$, the expression
for ${\cal L}_{tun}$ finally reads
\beq
{\cal L}_{tun} = 2^2\;\; \Gamma\;\; \cos\Big(\frac{1}{\sqrt{2g'}}
\phi(x=0,t)\Big)
\label{eq:cosphi}
\eeq

Once the chiral theory has been transformed into a non-chiral one,
we follow the procedure developed by Cardy to calculate the
scaling dimension. By transforming the two-point function of the
non-chiral field in a plane with boundaries, into a
four-point function of the chiral field in the entire
plane, we obtain:
\beq
\Delta_{FQH/NM}(\Gamma =0) = \frac{1}{g'} 
\eeq
for the  $1/3$ FQH junction, $g' = 1/2$ and
\beq
\Delta_{FQH/NM}(\Gamma =0) = 2
\eeq

Because the dimension of the boundary space is equal to 1, the value
of $\Delta_{FQH/NM}(\Gamma =0)$ implies that $\Gamma =0$ is an irrelevant
perturbation as expected.


\subsection{$\Gamma \to \infty$ Fixed Point}

To calculate the dimension of the tunneling term as appearing in the
dual picture of the junction, we proceed along the same lines
described above.  The dual tunneling Lagrangian now looks like
\beq
{\tilde {\cal L}}_{tun} = 2^2\;\; {\tilde \Gamma}
\cos\Big(\sqrt{\frac{g'}{2}} {\tilde \phi}(x=0,t)\Big)
\eeq

In this case we obtain the scaling dimension:
\beq
\Delta_{FQH/NM}(\Gamma \to \infty) = g' 
\eeq
and for the $1/3$ FQH junction: 
\beq
\Delta_{FQH/NM}(\Gamma \to \infty) = \frac{1}{2}
\eeq

Since $\Delta_{FQH/NM} <1$, the tunneling operator is a {\sl relevant}
perturbation at the strong coupling fixed point. Also, it was shown in
Ref.~\cite{ref:andreev} that, although exactly at the strong coupling
fixed point many scattering processes are forbidden, the main effect
of the tunneling operator is to increase the allowed scattering
processes that otherwise are suppressed. Hence, in this way the
presence of the tunneling operator drives the system from the
non-Fermi liquid to the Fermi-liquid fixed point. In analogy with
quantum impurity problems, it is useful to introduce an energy scale
$T_k$, associated with the impurity. At frequencies high compared with
the crossover scale $T_k$, the amplitude for these additional
scattering channels is exceedingly small and, in this regime,
non-Fermi liquid physics should be observable. However, this analysis
assumes that the {\sl only} relevant operator is the tunneling
operator and all other operators, such as multiparticle tunneling
processes, at the strong coupling fixed point are irrelevant.  While
at the weak coupling fixed point it is trivial to show that these
higher order processes are indeed strongly suppressed, it is not
obvious that this should be the case at the strong coupling fixed
point. Nevertheless it is reasonable to expect that there should exist
a regime in which these higher order processes can be effectively
ignored. In this case the model with just one relevant perturbation
makes physical sense.

Below we will show by comparison with the 2CK model, that this
tunneling operator has the same scaling dimension as the channel
symmetry breaking operator in the 2CK model. Intererstingly, it is
known that the channel symmetry breaking perturbation drives the 2CK
from the non-Fermi liquid to a Fermi liquid fixed point. Thus, it is
reasonable to conjecture that the tunneling operator in the FQH
junction corresponds to the channel symmetry breaking operator in the
2CK model. We will show below that, to an extent, this is true.

\subsection{Entropy at the non-Fermi liquid fixed point}
\label{sec:entropy}

The dual description describes the physics at strong coupling, {\it
i.e.}, the non-Fermi liquid fixed point. Thus, to obtain the entropy
at this fixed point it is necessary to work with the Lagrangian
describing the dynamics of the field ${\tilde \phi}_-$
(Eq.~(\ref{eq:duallag})) which corresponds to a boundary sine-Gordon
model. In the boundary sine-Gordon model, the interaction term gives a
boundary contribution to the energy (the boundary itself can hold
energy). As a consequence, energy conservation at the boundary implies
a dynamical boundary condition, {\it e.g.}, the boundary condition can
flow from Neumann $(\partial \phi(0)=0)$ to Dirichlet $(\phi(0)=0$).

In \cite{ref:FendleySW}, Fendley, Saleur and Warner (FSW) calculated
this boundary contribution to the free energy. They showed that in the
IR (low energy) limit, this contribution vanishes but it remains
finite in the UV (high energy) limit. From the free energy, they
obtained the boundary contribution to the entropy which, in the UV
limit, is given by $S={\frac{1}{2}} \ln (\lambda +1)$ where $\lambda$
is the compactification radius of the bosonic field:
$R={\sqrt{{\frac{\lambda+1}{2}}}}$.

By comparison with these results, we can obtain the value for the
entropy in the FQH/NM junction. In this case, the IR limit corresponds
to the $\Gamma = 0$ fixed point and the UV limit to the $\Gamma
\rightarrow \infty$ fixed point. Thus, we obtain a vanishing entropy
at $\Gamma = 0$ and a finite entropy at $\Gamma \rightarrow
\infty$. Since the compactification radius is
$\frac{1}{\sqrt{2g'}}=\sqrt{\frac{k+1}{2}}$, the value of the entropy
is given by $S={\frac{1}{2}} \ln (k+1)$. Furthermore, for the special
case of $k=1$, the value of the entropy is $S={\frac{1}{2}} \ln
(2)$. In other words, a FQH/NM junction with an effective $g'=1/2$
{\it i.e.}, a $\nu = 1/3$ junction, has a boundary contribution to the
entropy of $S={\frac{1}{2}} \ln (2)$ at the $\Gamma \rightarrow
\infty$$ ({\tilde \Gamma} = 0$). This is precisely the same value
found for the two-channel, spin-${\frac{1}{2}}$ Kondo model at the
non-trivial fixed point. Thus, we have proved that the $\nu = 1/3$
FQH/NM junction and the 2CK model have the same value for the entropy
at the non-Fermi liquid fixed point.


\section{Green's Function for electrons on the normal side of a FQH/NM junction}
\label{sec:greenfunction}

As shown in Ref.~\cite{ref:andreev}, the scattering processes allowed
in a junction between a normal metal and a FQH liquid are different
at the two different fixed points $\Gamma = 0$ and $\Gamma \ra
\infty$. While for $\Gamma = 0$ the scattering matrix corresponds to
a Fermi liquid fixed point, for $\Gamma \ra \infty$ the scattering
matrix is non-unitary. This non-unitarity is reflected by the
vanishing of the electron-electron scattering matrix element.
Usually, the manifestation of Fermi or non-Fermi liquid behavior
appears in the electron Green's function. In this section, we present
a standard perturbation theory calculation of the Green's function for
electrons on the normal metal side of the junction. In what follows
instead of working with the tunneling amplitud $\Gamma$, it is more
convenient to work with the energy scale $T_k$ introduced
earlier. This energy scale is related to the tunneling amplitud by
$T_k = 4\pi |\tilde{\Gamma}|^2$. Because $\tilde{\Gamma} = 0$ is and
unstable fixed point and the perturbation is a relevant operator (as
we show in Sec.~\ref{sec:dimensions}), it is important to remark that
the expression obtained for the Green's function is valid in the
regime of high energies or large momenta compared with $T_k$, {\it
i.e.}, when $T_k$ is the smallest energy scale. The electron's
propagator in real space-time is given by:
\begin{equation}
G_e(x,x',t,t') = -i \langle T [:e^{-i\phi_2(x,t)}:
:e^{i\phi_2(x',t')}:]\rangle
\label{eq:gerst}
\end{equation}
{\it i.e.} an electron is created at $(x', t')$ and destroyed, past
the point contact, at $(x,t)$. Here $T$ represents the time-ordered
product and $:\cdots :$ represents the normal ordered operator. Notice
that this definition presupposes that the electron propagator depends
on $x,x'$ and not necessarily on the difference $x-x'$, {\it i.e.},
translation invariance is broken due to the presence of the point
contact. The Fourier transform of this quantity is given by:
\begin{eqnarray}
& & G_e(\omega_o, k_o; \omega_i, k_i) = \int_{-L}^L dx dx'
\int_{-\infty}^{\infty} dt dt' e^{-i(k_ox-\omega_ot)} \nonumber \\ & &
e^{i(k_ix' - \omega_i t')} \;\;G_e(x,x',t,t')
\label{eq:greensfourier}
\end{eqnarray}

This expression can be interpreted as the propagator for an electron
created in the state $(k_i,\omega_i)$ and destroyed in the state
$(k_o,\omega_o)$. Because the presence of the impurity breaks
translation invariance, momentum is not a good quantum number for this
problem.  Nevertheless far away from the impurity site, the normal
metal is a Fermi liquid with translation invariance. If we consider
the scattering of electrons that are created and destroyed {\it
asymptotically} far away from the impurity, then we can use the
operator $:e^{i\phi_2(x,t)}:$ to represent electron states (or use
momentum eigenvalues to label electron states).

We can calculate the expression for $G_e(\omega_o, k_o; \omega_i,
k_i)$ at both fixed points. In particular, at the $\Gamma =0$ fixed
point, it is straightforward to show that the electron's Green's
function presents the characteristic pole structure that is the
signature of Fermi liquid behavior. To see this, notice that the mean
value in this case, is taken with respect to an action $S = S_1 +
S_2$, where $S_{1,2}$ are the actions of free chiral fermion fields.
Thus, we obtain:
\begin{equation}
G_e(\omega,k) \propto \frac{1}{\omega + k -i\epsilon}
\end{equation}
{\it i.e} the Green's function has the pole structure corresponding to
a Fermi liquid as expected. The details of the calculation are given
in Appendix~\ref{sec:app1}.

The tunneling term, represents a perturbation by an
irrelevant operator (see Sec.~\ref{sec:dimensions}), and as such, it
is not expected to introduce appreciable changes in the low-energy
physics of the system. In particular, this pole structure is
effectively the leading term in the expression for the Green's
function, when the tunneling perturbation is included.

To calculate $G_e(\omega_o, k_o; \omega_i, k_i)$ at the strong
coupling fixed point, we use Eq.~(\ref{eq:dualfi4}). Notice that in
this case, the perturbative expansion is around the non-Fermi liquid
fixed point (at $\tilde{\Gamma} = 0$), hence the mean value is taken
with respect to the action containing the dual Lagrangian given in
Eq.~(\ref{eq:duallag}). As stated above, this perturbative calculation
is legitimate only at high enough energies or temperatures; it can be
done in equilibrium conditions, where there is no external voltage
applied to the junction, and in non-equilibrium conditions. In what
follows we will describe first the equilibrium situation and later we
will show that non-equilibrium corresponds only to a shift in
frequencies for the incoming electron. We will restrict to the case
where $k_0 = k_i=k$, {\it i.e.}, the incoming and outgoing electrons
are in the same momentum state and by convenience we define
$\omega=\omega_i$. By standard perturbative techniques we calculate
$G_e(\omega,k)$ as follows
\begin{equation}
G_e(\omega,k)=-i\sum_{n=0}^{\infty}\frac{(-i)^n}{n!}\langle T (S_I)^n
e^{-i\phi_2(x,t)} e^{i\phi_2(x',t')}\rangle_{S_0({\tilde{\phi}}_{\pm})}
\label{eq:gepert}
\end{equation}
where 
\begin{equation}
S_I = \int_{-L}^L dx \int_{-\infty}^{\infty} dt \;\tilde{\Gamma}
\;\delta(x) \;\cos(\sqrt{2g'} \tilde{\phi}_-)
\end{equation}

Because the interacting action is given in terms of the dual fields,
it is necessary to write the field for the electron operator
$\phi_2(x,t)$ in terms of its duals. Thus, we work with the inverse
of the dual transformation:
\beqa
&& \lefteqn{\left(\matrix{\phi_1\cr \phi_2}\right)
=
\left[
\frac{\Theta(-x)}{\sqrt{2}} \left(\matrix{\cos\theta - 
\sin\theta & \cos\theta + \sin\theta\cr 
\cos\theta + \sin\theta & \sin\theta - \cos\theta}\right) \right.}
\nonumber \\ 
&+&
\left. 
\frac{\Theta(x)}{\sqrt{2}}
\left(\matrix{\cos\theta - \sin\theta & - \cos\theta - \sin\theta\cr
\cos\theta + \sin\theta & - \sin\theta + \cos\theta}\right)
\right] 
\left(\matrix{\tilde{\phi}_+\cr \tilde{\phi}_-}\right)
\nonumber \\ 
&&
\label{eq:dualrotation}
\eeqa

\subsection{Order n =0}

It is straightforward to show that at this order
\beq
G_e^{(0)}(x,x',t,t') = 0
\eeq

To see this result notice that when the field $\phi_2$ is
written in terms of the dual fields ${\tilde \phi}_{+,-}$, the mean
value factorizes into a product of two terms:
\beqa
&\langle& :e^{-i \sqrt{\frac{g'}{2\nu}} {\tilde \phi}_+(x,t)}: :e^{i
\sqrt{\frac{g'}{2\nu}} {\tilde \phi}_+(x',t')}: \rangle_{S_0({\tilde
\phi}_+)}
\nonumber \\ 
& \times&  \langle :e^{-i\sqrt{\frac{g'}{2}}{\tilde \phi}_-(x,t)}: 
:e^{-i \sqrt{\frac{g'}{2}}{\tilde \phi}_-(x',t')}:
\rangle_{S_0({\tilde \phi}_-)}
\nonumber
\\  &=& 0
\eeqa

In terms of a Coulomb gas interpretation, this is nothing else that
the consequence of the neutrality condition of the Coulomb gas.

\subsection{Order n=1}

By using the expansion in Eq.~(\ref{eq:gepert}) we find that, at first
order in ${\tilde \Gamma}$, the expression for $G_e(x,x',t,t')$ is
also given by a product of two factors:
\beqa
&& G_e^{(1)}={\tilde \Gamma}\int_{-\infty}^{\infty}dt_1 
\langle:e^{-i \sqrt{\frac{g'}{2\nu}} {\tilde \phi}_+(x,t)}::e^{i
\sqrt{\frac{g'}{2\nu}} {\tilde \phi}_+(x',t')}: \rangle_{S_0({\tilde \phi}_+)}
\nonumber \\ &&\langle:e^{-i\sqrt{\frac{g'}{2}}{\tilde\phi}_-(x,t)}: :e^{-i
\sqrt{\frac{g'}{2}}{\tilde\phi}_-(x',t')}:
:e^{i\sqrt{2g'}{\tilde\phi}_-(0,t_1)}:\rangle_{S_0({\tilde \phi}_-)}
\nonumber \\
&&
\label{eq:ge0}
\eeqa

Using the fact that
\beqa
&&\langle T:e^{i \alpha_1 \phi (x_1,t_1)}:\cdots 
:e^{i \alpha_n \phi (x_n,t_n)}:\rangle_{S_0(\phi)} = \nonumber \\
&& e^{- 2 \sum_{i<j}^n \alpha_i \alpha_j 
\langle T [\phi(x_i,t_i) \phi(x_j,t_j)]\rangle}
\label{eq:meanvalues}
\eeqa
where
\beqa
&&\langle T[\phi(x_i,t_i) \phi(x_j,t_j)]\rangle = \nonumber \\
&&-\ln[\delta+i {\rm sgn}(t_i-t_j)(t_i-t_j-x_i+x_j)]
\label{eq:chiralbosonprop}
\eeqa
we obtain the following expression for $G_e(k,\omega)$
\beqa
& &G_e={\tilde \Gamma} \int_{-\infty}^{\infty} d\omega_1dk_1 
~R_{\alpha}(\omega_1, k_1) ~~R_{\alpha}(-\omega_1,-k_1) \nonumber \\
& & ~~~~~~~~~~~\times~R_{1-\alpha}(\omega-\omega_1, k-k_1)+h.c. 
\label{eq:geeq} \\
&& R_{\alpha}(\omega, k) =  \int_{-\infty}^{\infty} dx\;dt\; 
\frac{e^{i (\omega t -kx)}}{[\delta + i {\rm sgn}(t)(t-x)]^\frac{2\nu}{1+\nu}}   \\    
&& R_{1-\alpha}(\omega, k) =  \int_{-\infty}^{\infty} dx\;dt\; 
\frac{e^{i (\omega t -kx)}}{[\delta + i {\rm sgn}(t)(t-x)]^\frac{1-\nu}{1+\nu}}
\eeqa
(where $\alpha = 2\nu/(1+\nu)$).  The explicit expression for
$R_{\alpha}(\omega, k)$ and $R_{1-\alpha}(\omega, k)$ are given in
Appendix~\ref{sec:app2}.
Finally, the expression for the Green's function depends on the value
of $\nu$ and it is given by:
\begin{itemize}
\item
For $\nu > 1/3$
\end{itemize}
\beqa
&&G_e^{(1)}(\omega,k) =  
{\tilde \Gamma} \;C(\nu) \;\frac{1}{(\omega - k + i\epsilon)^2}
k^{-(\frac{1-\nu}{1+\nu})}    \nonumber \\
&&
\label{eq:gemayor}
\eeqa
where
\begin{equation}
C(\nu) = 2^{(\frac{3\nu -1}{1 +\nu})} \;\frac{(2\pi)^4}{\sqrt{\pi}} 
\;\frac{\Gamma\left[\frac{1+5\nu}{2(1+\nu)}\right]}{\Gamma^2
\left[\frac{2\nu}{1+\nu}\right]} \left(\frac{1+\nu}{3\nu -1}\right) 
\label{eq:Cnu}
\end{equation}
Here $\Gamma(x)$ is the Gamma function and $\epsilon$ is an infrared
(IR) cutoff used in the calculation of the chiral fermion propagators
$R_{\alpha}$ and $R_{1-\alpha}$.

\begin{itemize}
\item
For $\nu =1/3$
\end{itemize}
\beqa
&&G_e^{(1)}(\omega,k) = \frac{(4\pi)^2}{(\omega - k + i\epsilon)^2}
\left(\frac{k}{\pi T_k}\right)^{-(\frac{1}{2})}
\ln\left[\frac{k}{T_k}\right]
\nonumber \\
&&
\label{eq:geigual}
\eeqa

Here we have used the fact that $T_k \propto {\tilde \Gamma}^2$. These results
hold only at large energies (and momenta) and thus $T_k$ plays the role of an IR
cutoff.
\begin{itemize}
\item
For $0 < \nu < 1/3$
\end{itemize}
\beqa
&&G_e^{(1)}(\omega,k) =  \frac{C''(\nu)}{T_k^{\sqrt{2}}}\;
\frac{k^{-(\frac{2\nu}{1+\nu})}}{(\omega - k + i\epsilon)^2} 
\Phi\left(\frac{3\nu -1}{1+\nu}; \frac{4\nu}{1+\nu}; \frac{T_k}{k}\right) 
\nonumber \\
&&
\label{eq:gemenor}
\eeqa
where
\begin{equation}
C''(\nu) =\;\frac{(2\pi)^4}{\Gamma^2 \left[\frac{2\nu}{1+\nu} \right] \;
\Gamma\left[\frac{1-\nu}{1+\nu}\right]} \;\left(\frac{1+\nu}{1-3\nu}\right)\; 
\frac{1}{T_k^{(\frac{1- 3\nu}{1+\nu})}}  
\end{equation}
where $\Phi(x,y,z)$ is the confluent hypergeometric function.

Our results for the Green function at the strong coupling fixed point
show that it has a branch cut singularity, in clear contrast with the
pole structure that appears at weak coupling. The case $\nu=1/3$ is
special in that it has a logarithmic branch cut.  Furthermore, the
structure of the expressions for the Fermion propagator has the form:
\beq
G_e^{(1)}(\omega,k) = G_{\circ}^{out} \;\Upsilon\; G_{\circ}^{in}
\eeq
where $G_{\circ}^{out,in}$ is the Green's function for a chiral Fermi
liquid. By inspection we see that $\Upsilon$ plays the role of a vertex function, and
as such it yields the leading order behavior for the one-body
S-matrix.  Notice that exactly {\sl at the strong coupling fixed
point} the one-body S-matrix is zero. The results of
Eqs.~(\ref{eq:gemayor}),(\ref{eq:geigual}) and (\ref{eq:gemenor})
represent the leading non-vanishing behavior close to the strong
coupling fixed point. These results apply at frequencies high compared
with the crossover scale $T_k$. (Notice that at this level the
frequency dependence enters only through $G_{\circ}^{out,in}$.)

The branch cut singularity of $\Upsilon$ and the vanishing of the
one-body S-matrix show that at the strong coupling fixed point
scattering is incoherent ({\it i.e.}, there is a broad continuum of
multiparticle scattering processes) which signals the breakdown of the
simple particle picture of the decoupled Fermi liquid. This result is
superficially reminiscent of scattering processes off a black hole,
with the broad continuum seemingly playing the role of Hawking
radiation.

An interesting point in this calculation is the dependence of $G_e$ on
the filling fraction $\nu$. This dependence can be understood in terms
of a {\it screening} process by using the Coulomb gas interpretation
for the expectation values. In terms of the field $\phi_2$ we see that
the dimensions of the operator representing the insertion of a charge
due to the impurity ($e^{-i g{\tilde \phi}_2(x=0,t_1)}$) and the
operator representing the asymptotically outgoing electron ($e^{-i
\phi_2(x,t)} \propto  e^{-i (1-g){\tilde \phi}_2(x,t)}$),  depend 
on $\nu$. In particular, for $\nu = 1/3$ these dimensions have the
same value of $1/2$. In terms of a Coulomb gas interpretation the unit
charge created at $(x',t')$ is partially or almost totally screened by
the inserted charge representing the effects of the impurity, while
the unscreened part of it is the charge seen at $(x,t)$.  For $\nu >
1/3$ the inserted charge is very effective in screening the charge
created at $(x',t')$ and thus, the charge seen at $(x,t)$ is small. As
$\nu$ decreases, the screening diminishes and, in particular at $\nu =
1/3$ the inserted charge screens it only by half. In other words, for
$\nu = 1/3$ these dimensions have the same value of $1/2$, as it was
observed previously in Ref.~\cite{ref:kane}. Finally, for $\nu < 1/3$
the screening by the inserted charge is almost completely ineffective.

\subsection{Non-equilibrium case}
\label{sec:nonequ}

Finally it is interesting to consider a non-equilibrium situation. For
this purpose we include a voltage difference between the reservoir and
the FQH edge by replacing the original coupling constant $\Gamma$ by
$\Gamma e^{-i\omega_0 t}$. Repeating the same procedure outlined
above, we find that Eq.~(\ref{eq:geeq}) is modified as follows:
\beqa 
G_e(k,\omega) &=& {\tilde \Gamma} \int_{-\infty}^{\infty} d\omega_1
\; dk_1 R_{\alpha}(\omega_1, k_1) R_{\alpha}(-\omega_1 + \omega_0, -k_1) \nonumber \\
              & & \times R_{1-\alpha}(\omega - \omega_1 + \omega_0, k
              - k_1) + h.c.
\eeqa

Carrying out the rest of the algebra we find that the effect of
the voltage is to change the Green's function of the outgoing electron
($G_{\circ}^{out}$) by 
\beq
\frac{1}{(\omega - k + i\epsilon)^2} \ra \frac{1}{(\omega - k + i\epsilon)} \frac{1}{(\omega + \omega_0 - k + i\epsilon)}
\eeq
Hence, to leading order, the external voltage does not affect th
branch cut
 singularity of the Green's function.
\section{Mapping of the FQH/NM junction to the 2CK problem: does it work?}
\label{sec:electron}

Given all the similarities between the $\nu = 1/3$ FQH junction and
the 2CK model, it is natural to look for an explicit mapping that can
relate not just the (formal) Hamiltonians, but also the operator
content of both models.  In particular, since in the FQH/NM junction
the strong coupling properties can be connected to the weak coupling
ones through a duality transformation, it is possible to look for a
mapping between the electron operator as defined in the FQH/NM
junction at couplings $\Gamma = 0$ and $\Gamma \to \infty$ . It is
also natural to ask whether the fermions of the reservoir side of the
junction are related in any way to the fermions of the 2CK
model. Since the model for the FQH/NM junction is written in the
bosonized language, it is useful to work with the 2CK model also in
the bosonized version.

We begin our mapping by showing that the (bosonized) Hamiltonian for
the FQH/NM junction (in its dual form) can be mapped onto the
(bosonized) Hamiltonian of the 2CK model, in the Tolouse limit.
Following Ref.~\cite{ref:Schofield}, the Hamiltonian for the 2CK model
after performing the Emery-Kivelson
transformation\cite{ref:emery-kivelson} reads:
\beqa
H_0' &=& H_0 \nonumber \\
H_{\perp}' &=& \frac{J^{(1)}_p}{4\pi \alpha} (-1)^{N_{c_+}  + N_{s_-}}
[S^+\; {e}^{-i \phi_{s_-}(0)} - h.c.] \nonumber \\
&+& \frac{J^{(2)}_p}{4\pi \alpha} (-1)^{N_{c_-}  - 
N_{s_-}} [ S^+\; {e}^{i \phi_{s_-}(0)} - h.c.] \nonumber \\
H_{\parallel} &=& (J_z - 2\pi v_F) \frac{S_z}{2\pi} \partial_x
\phi_{s_+}(0) + \delta J_z \frac{S_z}{2\pi} \partial_x \phi_{s_-}(0)
\label{eq:H2ck}
\eeqa
where $H_0$ is the spin-$1/2$ free fermion Hamiltonian, written in
terms of two chiral bosons one for charge and one for spin for each
channel denoted by $\pm$. Each chiral boson has compactification
radius (Luttinger parameter) equal to one. In Eq.(\ref{eq:H2ck}),
$J^{(i)}_p$ is the component of the coupling constant $J$ on the plane
for channel $i$, $J_z^{(i)} = (J^{(1)}_p +J^{(2)}_p)/2$ and 
$\delta J_z = (J^{(1)}_p - J^{(2)}_p)/2$. Also, since the charge modes decouple,
their Hamiltonian has been absorbed in $H_0$
\begin{figure}
\vspace{.2cm}
\noindent
\hspace{.8cm}
\epsfxsize=3.0in
\epsfbox{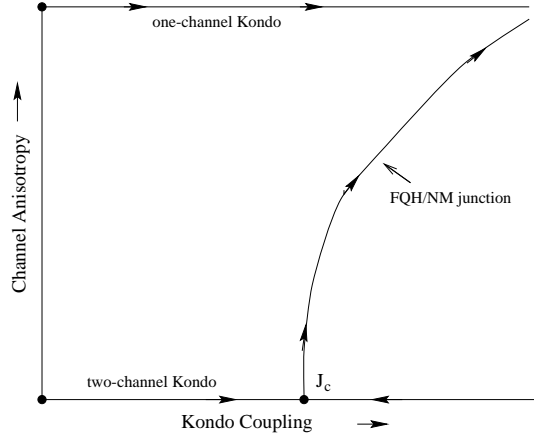}
\vspace{0.9cm}
\caption{Qualitative phase diagram of the two-channel 
Kondo model. The horizontal axis is the Kondo coupling $J$ and the
vertical axis is the channel anisotropy. $J_c$ is the infrared
non-trivial fixed point of the 2CK system. The FQH/NM junction is
mapped onto the RG trajectory that flows from the 2CK stable fixed
point to the (screened) stable fixed point of the 1CK system. The top
line, at infinte channel anisotropy, is the 1CK system. The arrows
show the qualitative RG flows.}
\label{fig:fig3.fig}
\end{figure}

In order to compare the Lagrangian for the 2CK model, with the Lagrangian for
the FQH/NM junction at $\nu=1/3$, we will rescale time and space so that
the constants $\alpha$ and $v_F$ take the values $\alpha = \frac{1}{4 \pi}$ and $v_F = 1$.
For simplicity we also take $J^{(2)}_p = 0$, {\it i.\ e.\/} we represent the
anisotropy in the channels by taking $J^{(1)}_p \neq 0$. Finally, following the approach of
Emery and Kivelson\cite{ref:emery-kivelson}, we
recognize that both Lagrangians have similar structures at the Toulouse point,
where $J_z = 2\pi v_F$ and $\delta J_z = 0$. It is well known that in the 2CK model the
Tolouse point flows under the RG to the isotropic 2CK model.

Within these conventions, the Lagrangian for the anisotropic 2-channel
Kondo model reads:
\beqa
&&{\cal L}_{2CK} = \frac{1}{4\pi} \int_{-\infty}^{\infty} \; dx\;
\partial_x \phi_{s_-} (\partial_t - \partial_x) \phi_{s_-} - \nonumber \\
&& J^{(1)}_p
\delta(x)\;\;(-1)^{N_{s_-}} [ S^+\; {e}^{-i \phi_{s_-}(x,t)} - h.c.] 
\label{eq:l2ck}
\eeqa

Here we have set $N_{c_+}=0$ because the field $\phi_{c_+}$ does not have
dynamics and $N_{s_-}$ represents the total charge of the field $\phi_{s_-}$.

The Lagrangian for the FQH/NM junction is:
\beqa
&& {\cal L}_{FQH} = \frac{1}{4\pi} \int_{-\infty}^{\infty} \; dx \;\partial_x
{\tilde \phi}_- (\partial_t - \partial_x) {\tilde \phi}_- + \nonumber \\
&& 2 \;{\tilde\Gamma}\; \delta(x)\; 
{\eta}^{\dagger}_{edge} \eta_{res} [{e}^{-i{\tilde \phi}_-(x,t)} + {e}^{i{\tilde \phi}_-(x,t)}]
\label{eq:lfqhj}
\eeqa

In these expressions we have explicitely left out the Lagrangian corresponding
to the free field ${\tilde \phi}_+$ that decouples from the impurity, after
making the appropriate transformations.

Comparing Eqs.~(\ref{eq:l2ck}) and (\ref{eq:lfqhj}), we make the
identifications:
\beqa
J^{(1)}_p & \to & 2 {\tilde \Gamma} \nonumber \\
(-1)^{N_{s_-}} & \to & {\eta}^{\dagger}_{edge} \eta_{res} 
\eeqa

Clearly, the following identification between operators can be also
made:
\beqa
{e}^{-i{\tilde \phi}_-(x,t)} & \equiv S^+\; {e}^{-i \phi_{s_-}(x,t)}
\nonumber \\
{e}^{i{\tilde \phi}_-(x,t)} & \equiv S^-\; {e}^{i \phi_{s_-}(x,t) + \pi} 
\label{eq:mapping}
\eeqa

From these expressions we see that the dual field $\tilde\phi_-$ is
given by a composite operator involving spin and fermion operators of
the 2CK. Thus, although the Hamiltonian can be mapped into each other,
the degrees of freedom are glued in a particular way. In other words,
what in one system has a simple expression it becomes a more
complicated object in the other.

These expressions show that, at the level of their Hamiltonians the
2CK model and the FQH/NM junction at $\nu=1/3$ apparently do map into each
other. Notice that in reality we have not mapped the junction into the
full 2CK model but only to the Tolouse limit of this model. However,
since the full 2CK model flows (under RG) to the Tolouse limit (under
the RG), then at this level the mapping is established. In particular,
the tunneling operator of the junction maps onto the channel
anisotropy term in the 2CK. Hence the RG flow of the FQH junction, which
is an integrable system\cite{ref:FendleySW}, maps onto the RG
trajectory of the 2CK model from the non-trivial 2CK fixed point to the
stable fixed point of the 1CK model. This trajectory has been studied
in detail by Andrei and Jerez\cite{ref:andrei}.

However, although these arguments are strongly suggestive, they are not by
any means sufficient to establish a mapping. In addition to a mapping
between the Hamiltonians, it is necessary to specify the Hilbert
spaces on which these Hamiltonians act and how these spaces may (or
may not) be related. In other terms, it is necessary to specify what
operators are physically allowed in each problem and how they may
possibly be related to each other. We will see, however, that although
the same bosonized Hamiltonian does describe both systems, the way the
physical operators are put together is quite different and that
operators that are physical in one system are not physical in the
other. Furthemore, even though the Hamiltonains are the same, we will
also find that the fields obey different boundary conditions.

Let us begin by seeking an expression for the electron operator on the
reservoir (the Fermi liquid side of the junction ), as defined at the
$\Gamma = 0$ fixed point, in terms of the field $\phi_{s_-}$. We can
do that by using the transformations between the fields $\phi_2$ and
$\tilde{\phi_-}$, (Eq.~(\ref{eq:dualrotation})) and we find,
\beqa
{e}^{i \phi_2(x<0)} &=& {e}^{i \frac{\sqrt{3}}{2} {\tilde \phi}_+ (x <0)}
{e}^{-i \frac{1}{2}{\tilde \phi}_-(x<0)} \nonumber \\
{e}^{i \phi_2(x>0)} &=& {e}^{i \frac{\sqrt{3}}{2} {\tilde \phi}_+ (x >0)}
{e}^{i \frac{1}{2}{\tilde \phi}_-(x>0)}
\label{eq:electron}
\eeqa

By comparison with Eq.(\ref{eq:mapping}), Eq.(\ref{eq:electron})
shows that the electron operator for the reservoir, $e^{i \phi_2}$,
cannot be put in terms of the spin and fermion operators that appear
in the two channel Kondo model. Notice that in the mapping of
Eq.(\ref{eq:electron}) the field ${\tilde \phi}_+$
enters. Furthermore, the operators that appear on the r.h.s. of
Eq. (\ref{eq:electron}) are not allowed in the Hilbert space of the
2CK model since they violate the required periodicity
conditions. Thus, the operator $e^{i \phi_2}$ is not contained in
the spectrum of the 2CK model. This is most remarkable since each
system contains an electron operator whose Green's function becomes
non-trivial at the respective non-trivial fixed point. Nevertheless
that by itself does not imply a relation and indeed the mapping
between the electron operators does not exist.

\begin{figure}
\vspace{.2cm}
\noindent
\hspace{0.4cm}
\epsfxsize=3.0in
\epsfbox{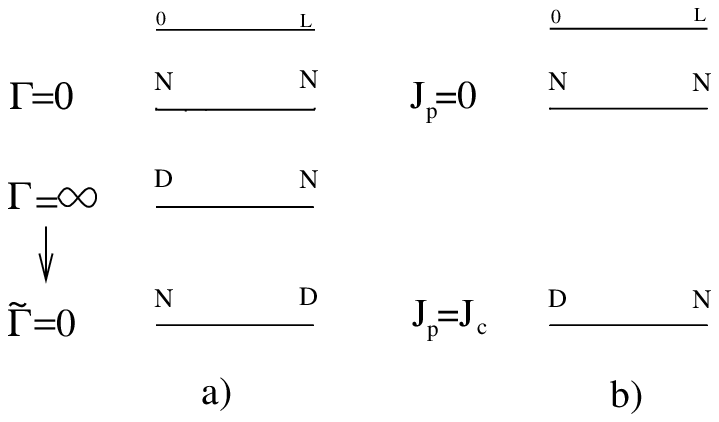}
\vspace{0.9cm}
\caption{Change in the boundary conditions following the flow of: a)
$\Gamma$ for the FQH junction (field $\tilde \phi_-$), and b) the
coupling $J$ for the 2CK model (field $\phi_{s_-}$).}
\label{fig:fig4.fig}
\end{figure}

Another way to understand the difference between the Hilbert spaces of
the two models, is to look at what boundary conditions do their fields
satisfy.  At the $\Gamma = 0$ fixed point the field $\phi_-$ satisfies
Neumann boundary conditions, {\it i.\ e.\/} $\partial \phi = 0$, both
at the point contact and at infinity. This condition follows from
current conservation: no current flows in or out of the system. When
the tunneling perturbation is applied, $\Gamma$ is $\neq 0$ and, as
discussed in Sec.~\ref{sec:entropy}, it induces a flow in the boundary
condition at the point contact. In particular, as $\Gamma$ goes to
$\infty$, the boundary condition becomes Dirichlet: $\phi_-=0$. This
happens because at $\Gamma \rightarrow \infty$ there is a perfect
contact between the two systems and $\phi_1(x=0) = \phi_2(x=0)$ giving
$\phi_-=0$. To study the strong coupling physics the duality
transformation is performed. It is also well known that under a
duality transformation Dirichlet and Neumann boundary conditions are
exchanged.  Thus, the boundary conditions on the dual field
$\tilde{\phi_-}$ at the strong coupling fixed point are now {\sl
Neumann at the origin} and {\sl Dirichlet at infinity}. The process of
changing boundary conditions is shown in Fig.\ref{fig:fig4.fig}(a).

Now, let us analyze the boundary conditions on the field $\phi_{s_-}$
of the 2CK model. At the trivial fixed point the boundary conditions
are Neumann, both at the point contact and at infinity. Turning on the
perturbation with the impurity induces a flow of boundary conditions
just as before. When the perfectly symmetric non-trivial fixed point
is reached the flow has changed from Neumann to Dirichlet at the
origin. Thus, in the non-trivial fixed point the boundary conditions
for the field $\phi_{s_-}$ are {\sl Dirichlet at the point contact} and
{\sl Neumann at infinity}. Hence, the boundary conditions at the non-Fermi
liquid fixed point of the 2CK are exactly the reverse of the ones of
the FQH/NM junction!. The logical implication is that, even though
both models have several common properties, their operator content is
actually quite different. In particular, physical quantities
associated with the electron operators in both systems are not simply
related to each other.

\section{Conclusions}
\label{sec:conclusions}

In this paper we have analyzed in detail the nature of the relation
between FQH/NM junctions and the 2CK model. We have found an explicit
mapping between their respective Hamiltonians and shown how to
construct the operators of interest.  We have also shown that both
models possess the same value for the residual entropy in the strong
coupling regime by using exact results derived for the boundary
sine-Gordon model. Furthermore, by performing a perturbative
calculation at the strong coupling fixed point, we have found a branch
cut singularity in the Green's function for electrons on the normal
side of the junction, making manifest the non-Fermi liquid nature of
this fixed point.  We also showed that the scaling dimensions and the
RG flows followed by the tunneling perturbation in the junction
suggest that this tunneling operator could be mapped into the channel
symmetry breaking perturbation in the 2CK model. In both cases the
dimension is $1/2$ and the flow goes from the non-Fermi liquid to the
Fermi liquid fixed point. Nevrtheless, we have also shown that in
spite of this closely related behaviors, the two problems are not
actually equivalent to each other.  We found instead that both models
do not share the same operator spectrum.  In order to understand the
difference we analyze the boundary conditions at the strong coupling
fixed point. We showed that while the FQH junction satisfies Neuman
at the point contact and Dirichlet at infinity, the boundary
conditions for the 2CK model are the opposite. Thus, we conclude that
although the two models share many common physical properties, their
Hilbert spaces cannot be mapped into each other.

\section{Acknowledgements}
\label{sec:ack}

NS acknowledges Hubert Saleur for helpful discussions. We are grateful
to Natan Andrei, Claudio Chamon and Andreas Ludwig for many
discussions and suggestions. During the completion of this work EF was
a participant of the Program on High Tc Superconductivity at the
Institute for Theoretical Physics, UCSB.  This work is supported in
part by NSF grants DMR-99-84471 at Brandeis University (NS),
DMR-98-17941 at UIUC (EF) and PHY94-07194 at ITP-UCSB (EF).

\appendix

\section{Electron Propagator at the $\Gamma =0$ Fixed Point.}
\label{sec:app1}

In this appendix we give the details of the calculation for the electron
Green's function at the $\Gamma =0$ fixed point.

The definition of the electron creation operator is given by
\beq
\psi^{\dagger}(t,\sigma) = \eta :e^{-\frac{i}{\sqrt{\nu}} \phi_L(t,\sigma)}:
\eeq
where the definition of the normal ordered exponential is given by
\cite{ref:Green}
\beqa
&& :e^{-\frac{i}{\sqrt{\nu}} \phi_L(t,\sigma)}: =
e^{-\frac{i}{\sqrt{\nu}} [\phi_0 +p_0\;(t + \sigma)+ i
\;\sum_{n=1}^{\infty} \frac{1}{n} \alpha^{\dagger}_n
e^{in(t+\sigma)}]} \times \nonumber \\ &&
e^{\frac{i}{\sqrt{\nu}}[\sum_{n=1}^{\infty} \frac{i}{n}
\alpha_ne^{-in(t+\sigma)}]}
\eeqa

Notice that for the electron on the normal metal side of the
junction $\nu =1$.  To calculate the electron's propagator we need to
evaluate the expression
\beq
G_e(t,\sigma) = -i\;\;\Big\langle 0 \Big| T \Big[ :\psi(t,\sigma)
\psi^{\dagger}(0,0): \Big] \Big| 0 \Big\rangle
\label{eq:geapp2}
\eeq
that is, we need to calculate products of the type
\beq
:\psi(t,\sigma) \psi^{\dagger}(0,0): \;\;\;;\;\;\;
:\psi^{\dagger}(0,0)\psi(t,\sigma):
\eeq

To normal order the products involving exponentials we make use of the
identity
\beq
e^A\;e^B = e^{[A,B]}\; e^B\;e^A
\eeq

A typical commutator to evaluate looks like
\beq
[A,B]=\Big(\frac{-1}{\nu}\Big)\sum_{n=1}^{\infty} \frac{-i}{n}
e^{-in(t+ \sigma)} \sum_{m=1}^{\infty} \frac{i}{m} [\alpha_n,
\alpha_m^{\dagger}]
\eeq

After  all the steps to normal order these products are carried out, we
have to calculate expressions like the following one:
\beq
\sum_{m=1}^{\infty} \frac{1}{m} e^{-im(t+ \sigma)}
\eeq
that we evaluate using an analytic continuation on time $t \to t(1-i\epsilon)$
with $\epsilon >0$. Thus, we find that 
\beq
\sum_{m=1}^{\infty} \frac{1}{m} e^{-im(t+ \sigma)} = \ln\Big(\frac{1}{1 - e^{-i(t+ \sigma)}}\Big)
\eeq

Finally, we get the expressions:
\beqa
\langle :\psi(t,\sigma) \psi^{\dagger}(0,0):\rangle &=& 
\eta^2 e^{-i\frac{1}{2\nu}(t+\sigma )} \Big[1 - e^{-i(t+ \sigma)}\Big]^{-\frac{1}{\nu}} 
\nonumber \\
&& \;\;\;\; {\rm for} \;\;\;\;\; t > 0
\nonumber \\
\langle :\psi^{\dagger}(0,0) \psi(t,\sigma):\rangle &=& \eta^2 e^{i\frac{1}{2\nu}(t+\sigma )} \Big[1 - 
e^{i(t+ \sigma)}\Big]^{-\frac{1}{\nu}}
\nonumber \\
&& \;\;\;\; {\rm for} \;\;\;\;\;  t < 0
\nonumber \\
&&
\eeqa

Replacing these expressions in Eq.~(\ref{eq:geapp2}), we obtain
\beq
G_e(t,\sigma)=-i\frac{\Theta(t) \eta^2 e^{-i\frac{1}{2\nu}(t+\sigma
)}}{\Big[1- e^{-i(t+ \sigma)}\Big]^{\frac{1}{\nu}}} -i
\frac{\Theta(-t) \eta^2 e^{i\frac{1}{2\nu}(t+\sigma)}}{\Big[1-e^{i(t+\sigma)}\Big]^{\frac{1}{\nu}}}
\eeq

In the thermodynamic limit $(t, \sigma) \ll 1$ (in units of $L = 2\pi$), and
the expression for the electron Green's function can be approximated by
\beq
G_e(t,\sigma) = -i\eta^2 \Big[
\frac{(-i)}{t+\sigma}\Big]^{\frac{1}{\nu}} \Theta(t) -i\eta^2 \Big[
\frac{(i)}{t+\sigma}\Big]^{\frac{1}{\nu}} \Theta(-t)
\label{eq:gewen}
\eeq

Using the fact that on the normal metal side of the junction $\nu =1$,
and by Fourier transforming the expression given in
Eq.~(\ref{eq:gewen}) we finally obtain:
\beq
G_e(\omega,k)=2\pi\;\eta^2 \lim_{\delta \to
0}\Big[\frac{\Theta(k)}{\omega+k
-i\delta}+\frac{\Theta(-k)}{\omega+k+i\delta}\Big]
\eeq
as corresponds to a chiral Fermi liquid.

\section{Calculation of the functions $R_{\alpha}(\omega, k)$ and
$R_{1-\alpha}(\omega, k)$}
\label{sec:app2}

In this appendix we give the details of the calculation for the
expressions of the functions $R_{\alpha}(\omega, k)$ and $R_{1-\alpha}(\omega, k)$.
We represent both functions by a generic function $Q_{\alpha}(\omega, k)$
where
\beq
\alpha = \frac{2\nu}{1+\nu} ;\;\;\;\;\;\;\;\ 1-\alpha = \frac{1-\nu}{1+\nu}
\eeq

Thus,
\beq
Q_{\alpha}(\omega, k)=\int_{-\infty}^{\infty}dx dt 
\frac{e^{i\omega t} e^{-ikx}}{\Big[ \delta+i {\rm sgn}(t)(t-x)\Big]^{\alpha}}
\eeq

Because of the function ${\rm sg}(t)$ in the denominator, we need to evaluate
integrals like
\beq
\int_{-\infty}^{\infty}\;dx\; \frac{e^{-ikx}}{(\beta - ix)^{\alpha}} = 
\frac{2\pi k^{\alpha - 1}}{\Gamma(\alpha)} e^{-\beta k} \Theta(k)
\eeq
and
\beq
\int_{-\infty}^{\infty}\;dx\; \frac{e^{-ikx}}{(\beta + ix)^{\alpha}} = 
\frac{2\pi (-k)^{\alpha - 1}}{\Gamma(\alpha)} e^{\beta k} \Theta(-k)
\eeq
where $\beta = (\delta + i t)$.

By integrating these expressions with respect to the variable $t$ using an IR
cutoff $\mu$, we obtain
\beq
Q_{\alpha}=\frac{2\pi}{\Gamma(\alpha)}\Big[ \frac{\Theta(k) k^{\alpha
-1} e^{-\delta k}}{(\omega - k + i\mu)} +
\frac{\Theta(-k) (-k)^{\alpha -1} e^{\delta k}}{(\omega - k - i\mu)}\Big]
\eeq

By replacing the values of $\alpha$ by their corresponding functions of $\nu$
we obtain the expressions mentioned in Sec.~\ref{sec:electron}.


\end{multicols}
\end{document}